\documentstyle[aps,prl]{revtex}
\def\gapp{\mathrel{\raise.3ex\hbox{$>$}\mkern-14mu
              \lower0.6ex\hbox{$\sim$}}}
\def\gsim{\gapp}
\def\lapp{\mathrel{\raise.3ex\hbox{$<$}\mkern-14mu
              \lower0.6ex\hbox{$\sim$}}}
\def\lsim{\lapp}
\begin{document}
\draft
\preprint{DAMTP-R98/}
\title{Of Matter Less Repulsive than a Cosmological Constant}
\author{ Neil J. Cornish$^{\dagger}$, and Glenn Starkman$^{\ddag}$}
\address{$^{\dagger}$Relativity Group, DAMTP, Cambridge University,
Silver Street, Cambridge CB3 9EW, England}
\address{$^{\ddag}$Department of Physics, Case Western Reserve
University, Cleveland, Ohio 44106-7079}
\twocolumn[
\maketitle
\widetext
\begin{abstract}
The case grows ever stronger that 
the average density of matter, ordinary and dark,
is less than the critical density required for a flat universe.
However, most of determinations of the mass density have been dynamical,
hence sensitive only to matter which is clustered 
at or below the scale of the observed dynamical systems.
The density may still be critical if there is a dark matter component
which is relatively smooth on the scales of galaxies or clusters.
Thoughts on this matter have focused on the possibility of an
effective cosmological constant or vacuum energy.
In this letter we examine an alternative possibility - that there is
a second component to the dark matter which has a repulsive
self-interaction. We show that given even very weak self-repulsion,
this dark matter would remain unclustered. While this repulsive
alternative is perhaps aptly named, it is arguably at least as palatable
as a cosmological constant. 

\end{abstract}
]

\begin{picture}(0,0)
\put(400,190){{\large CWRU-P98/16}}
\end{picture}

\narrowtext


For many years theoretical cosmologists have favoured flat
models with critical total energy density $\Omega = 1$,
where $\Omega$ is dominated by non-relativistic matter.
This theoretical prejudice is based on the Dicke coincidence argument
and standard models of inflation. The simplest, most generic
models of inflation predict $\vert\Omega-1\vert\ll1$. Indeed,
until the observational evidence favouring low $\Omega$ in clustered
matter became overwhelming, $\vert\Omega-1\vert\ll1$ was widely viewed
as the unalterable, defining prediction of inflation (and still is by some).

Confronted by the growing evidence that clustered matter
does not add up to more than about 30\% of the critical density, 
two alternative interpretations have been widely circulated.
The first view is that the total energy density of the
universe is indeed subcritical, so the spatial geometry is hyperbolic,
not Euclidean. Advocates of this view relinquish the standard prediction, 
and standard models of inflation. The alternative view\cite{krauss}
is that the universe is indeed flat, but that the 70\% of the critical density
is in the form of a cosmological constant, also known as vacuum energy.  
Since vacuum energy does not cluster, such energy density or its absence,
would not have been detected by observations of the dynamics of
clustered systems. 

The main theoretical argument for a cosmological constant solution is
that the universe remains flat, ($\Omega=1$), so one can
continue to accept the standard inflationary models and their
predictions.  
Recent results on the Hubble parameter as a function of
redshift from observations of type Ia Supernovae\cite{SN},
may favour a cosmological-constant.dominated universe
over either an open universe or a flat matter-dominated universe.
There is also some preliminary evidence for a 
doppler peak in the cosmic microwave background power spectrum
at multipole number $\ell \sim 200$. This would rule out an open
universe with $\Omega = 0.3$, and make a strong case for a flat
universe\cite{bart}. But an $\Omega=1$ universe with $\Omega_o=0.3$ in
clustered matter does not prove the case for cosmological
constant. We need to consider every alternative possibility,
and put each of them to the test. Then, ``...when you have eliminated the
impossible, whatever remains, however improbable, must be the truth.''\cite{Sherlock}

One interesting alternative\cite{rob} is that there is an additional form of
unclustered dark matter, call ``quintessence'' or $Q$-matter with
equation of state $\rho=\omega p$ where $-1<\omega<0$. Since
$\omega<0$ this $Q$-matter exerts a repulsive tension that inhibits
clustering. The special case $\omega=-1/3$
was first considered by Kolb\cite{rocky} (Kolb dark matter), who
showed that one could then have a closed universe with $\Omega
>1$ even if the energy density in ordinary matter is
subcritical, $\Omega_o<1$. However, even this wide class of $Q$-matter
models does not exhaust the possibilities. Here we suggest another. 

It is possible that the universe is flat, 
but that 70\% of the critical density is composed 
not of vacuum energy but of unclustered non-relativistic matter.
The standard objection to this scenario is that
any ordinary matter will fall into the gravitational potential wells  
defined by galaxies, clusters and large scale structure,
and become {\it ipso facto}~ clustered matter.  
Our solution to this conundrum is to imbue the new type of matter,
which we term X-matter, with a repulsive self-interaction strong
enough to prevent clustering in existing non-linear structures.

Let us be the first to admit that such a repulsive alternative is
neither elegant nor natural. But, as well shall demonstrate,
current observational and theoretical constraints do not rule out the
possibility. So for now we must add  X-matter to our list of
possible improbables.

A simple way to accomodate repulsive matter is to postulate a new
broken gauge interaction under which ordinary matter, including ordinary dark
matter, is neutral (or more generally is a singlet), but under which the X
particle carries a charge. We will consider only the Abelian case 
and take the charge of the X to be unity.
The strength of the $U(1)_X$ gauge interaction is characterised 
by the analogue $\alpha_X$ of the fine structure constant,
with an implicit condition that $\alpha_X<1$. The range of the
interaction is determined by the mass $m_B$ acquired by the $U(1)$
gauge boson in the symmetry breaking process.
Because the gauge interaction is broken, 
there is no constraint on the total X-charge of the universe.
It has been suggested to us\cite{taylor} that
the $U(1)$ gauge interaction might somehow be connected with a gauging of
lepton and/or baryon number. We have not carefully
considered the complications and constraints that implies,
given the fact that baryon and lepton number
are carried by ordinary matter.

In the neighbourhood of a galaxy or other gravitational potential well,
a smooth background of X particles of density $n_X$ will develop
an overdensity ${\delta n_X / n_X}$.  
This overdensity will grow until the Coulomb repulsion of the X's
overwhelms their gravitational attraction to the ordinary matter.
The gravitational force on an X particle due to the galaxy or cluster is 
\begin{equation}
F_g = {G M m_X \over r^2}
\end{equation}
where $M$ and $r$ are the mass and radius of the galaxy or cluster.
The repulsive force on the X particle trying to fall into the galaxy
once there is an X  overdensity $\delta n_X$ is approximately:
\begin{equation}
F_c 
= {4\pi\alpha_X \hbar^2 \delta n_X \over 3 m_B }
\end{equation}

If $F_c>F_g$, then the X particle will not fall into the galaxy,
and so the overdensity of $X$'s will build until
\begin{equation}
{\alpha_X \delta n_X \over m_X m_B} \simeq
{3 G M  \over 4\pi \hbar^2 r^2} .
\end{equation}
Taking the astronomical system to be approximately spherical,
and expressing its mass  
in terms of its radius and average density $\rho_{\rm av}$, 
we can find the X overdensity:
\begin{equation}
\delta n_X = { G r \rho_{\rm av} m_X m_B \over \alpha_X \hbar^2} .
\end{equation}
Since the whole point of this exercise is to insist
that  the X matter not be clustered, we require 
that 
\begin{equation}
m_X \delta n_x \ll  \rho_{\rm av}
\end{equation}
This implies that 
\begin{equation}
{m_X^2 m_B \over \alpha_X } \ll
{\hbar^2 \over G r } = M_{Pl}^2 {\hbar \over r c}
\label{eqn:nocluster}
\end{equation}

For a cluster , a rather generous $r\simeq 10 Mpc$, gives 
\begin{equation}
{m_X^2 m_B \over \alpha_X } \ll 10^4 GeV^3/c^6
\label{eqn:clusternocluster}
\end{equation}
We  would also probably prefer $m_X, m_B \gg 1$eV in order
that the $X$'s not be  free streaming during the
growth of large scale structure.
These limits imply that X and B masses in the eV-GeV range are realistic.

Our third constraint comes from Standard Big Bang Nucleosynthesis (SBBN).
The success of the SBBN model in reproducing the light element abundances,
informs us that we cannot add too many light degrees of freedom.
In this instance, ``light'' means $m\lsim 1-10MeV$. Given our
constraint (\ref{eqn:clusternocluster}) above, 
there would be no problem taking  both $m_X$ and $m_B$ heavy enough
to evade this bound, however, we might also be interested
in having them lighter.
The maximum allowed is generally quoted to be approximately
one-third of a light neutrino family equivalent.  
This would be impossible if the light $X$ or $B$ was at the
same temperature as the ambient ordinary matter.
However, even ordinary neutrinos are expected to be colder
than the photons during SBBN.  

Neutrinos are  thermally coupled to  the photonos only
via weak interactions. Since the strength of the
weak interactions  falls rapidly with the energy of the constituents,
the neutrinos thermally decouple from the ambient plasma
at a few times $10^{10}$ $^o$K. When the electrons and positrons annihilate
soon after, their entropy is injected nearly entirely into the 
electro-magnetically  coupled photon-electron-nucleon plasma,
leaving the neutrinos slightly colder: 
$T_\nu = (4/11)^{1/3}T_\gamma$.
Which  is the cube root of the ratio of the effective number of relativitic
degrees of freedom after $e^+e^-$ annihilation,
to the effective number of relativistic
degrees of freedom before $e^+e^-$ annihilation.

An analagous calculation of the temperature of the $B$
finds that 

\begin{equation}
{T_B\over T_\nu} = \left({g_\nu \over g_B}\right)^{1/3} \, .
\end{equation}
Here $g_\nu=\frac{43}{4}$ is the effective number of relativistic
degrees of freedom before neutrinos thermally decouple.
(photons contribute $2$, electrons contribute $4\times \frac{7}{8}$,
neutrinos contribute $2\times \frac{7}{8}$ per family.)
Since 
\begin{equation}
{\rho_B\over \rho_\nu} =  {T_B^4 \over {7\over 8} (T_\nu)^4}
\end{equation} 
the requirement that the B contribute less than $0.3$ effective
neutrino families is $(g_\nu/g_B)^{4/3}<0.3\times \frac{7}{8}$.
This implies
\begin{equation}
g_B > 30
\end{equation}
This is easily accomplished;
in the $SU(3)\times SU(2)\times U(1)$ standard model 
$g \gsim 60$ before the $QCD$ phase transition at $T\simeq 150$MeV,
and $g >100$ at $T\gsim 100$GeV.
Since, the $B$ boson couples only to $X$ charge, 
and since ordinary matter is $X$-neutral,
the B will thermally decouple from the 
ordinary plasma as soon as the $X$ does.

This also indicates that within the standard model
it would also not be particularly difficult to accommodate
a light X ($m_X \lsim 10$MeV) in addition to a light B.
The limit would be  $g_B > 47$, if the X were a Weyl
fermion (like a standard left handed neutrino)
$g_B > 63.$ if the X were a Dirac fermion (like an electron)
and $g_B > 49.$ if the X is a boson.
Thus so long as the $X$ and B are $SU(3)_c\times SU(2)_L\times U(1)_Y$
singlets, they would have thermally decoupled early enough
to evade the nucleosynthesis constraints on 
light particles. Since any particle dark matter candidate
must have a small interaction cross-section with
ordinary matter\cite{strongdm}, the assumption that it
carries no $3-2-1$ quantum numbers is the simplest one.

We also want the $X$ particles to contribute $\Omega_X \simeq 0.7-0.8$
to the mass density of the universe.
This means that
\begin{equation}
n_X = {\Omega_X \rho_{\rm crit}\over m_X} \, ,
\end{equation}
where $\rho_{crit}=1.88\times10^{-26} h^2 \, kg \, m^{-3}$ is the critical
density, and $h$ is the Hubble constant in units of $100 km\, sec^{-1}
Mpc^{-1}$. The number of ${\bar X}$'s must be negligible compared
to the number of $X$'s (in order for Coulomb repulsion to 
keep X's from collecting in galaxies and clusters).
Thus there must be a nearly perfect $X-{\bar X}$ asymmetry.
Large asymmetries are, in general, difficult to 
produce in simple particle physics models.
The magnitude of the asymmetry can best be
characterised by the number density $n_X$ of $X$'s,
divided by the number density of bosons in thermodynamic equilibrium
at temperature $T_B$.
We thus require 
\begin{equation}
n_X  \ll {1\over 4}  T_B^3 .
\end{equation}
Expressing $n_X$ in terms of 
$m_X$, $\Omega_X$ and $\rho_{\rm crit}$;
and $T_B$ in terms of 
$g_B/g_\gamma$ and the measured value of $T_\gamma=2.72^o$K,
we find
\begin{equation}
m_X \gg 4.6\times10^{-7}h^2 GeV {\Omega_X\over 0.7} {g_B\over 100} 
\end{equation}
Indicating that this is {\em not} a particularly restrictive constraint.
For an $X$ mass of $1$GeV, we see that an $X-{\bar X}$ asymmetry
of only $\sim 3\times10^{-7}$ is required.

What are the difficulties with this scenario?
First there is the age problem. The age of an $\Omega=1$
matter dominated universe is $t_o=6.52\times10^9h^{-1}$yrs.
The current best estimate for the age of the oldest
globular cluster is \cite{chaboyer}, $11.5\pm1.3\times 10^9$yrs,
with a one-sided 95\% confidence level lower limit of $9.5$Gyr.
For $\Omega=1$, this implies $h \leq 0.67$, with no allowances
for any protracted time interval between matter domination
and globular cluster formation. An expansion rate of
$H_o= 67\, km\, sec^{-1} Mpc^{-1}$ is consistent with the lower end of
most recent determinations of the the Hubble parameter.

The second concern is the limits placed by observations
of the light curves of type Ia supernovae\cite{garanavich,riess}
on $\Omega_{nr}$ in non-relativistic matter, which exclude
$\Omega_{nr}=1$ at better than $95\%$ confidence level.
Nevertheless, there are still concerns over both the 
small sample size in these ongoing studies, and
over the systematic effects of reddening.

Finally, there are certainly questions of the ``naturalness''
of this model.  The fact that one seems to require
yet another form of dark matter with a considerable 
asymmetry is not particularly appealing.  On the other
hand, since the X-matter is non-relativistic, 
the ratio $\Omega_X/\Omega_B$ (and by implication
$\Omega_X/\Omega_{dm}$) would be time-independent
at temperatures below  the mass of the lightest
of the proton, the $X$ particle and the ordinary
dark matter particle. Thus $\Omega_{nr}=0.3$
would not define a special epoch. This is a philosophical motivation
not shared by either a hyperbolic universe or by a universe dominated
by a cosmological-constant.

The possibility of preserving the standard prediction of
inflation that $\Omega=1$, without resorting to a cosmological
constant holds some appeal.  The classic argument
forbidding this is that non-relativistic matter would clump
in existing gravitationally bound structures such as galaxies
and clusters to an extent not consistent with observations.
We have pointed out that this argument
relies implicitly on the argument that all forms of dark matter
are not self-interacting, and that a relatively generic
form of interaction could lead to non-clustering dark matter.
We make no attempt to extol the great beauty of this model, 
nor to identify the X-matter with any particular
well-motivated, technically or philosophically natural
particle physics candidate, leaving this
rather to the reader's own imagination.

\end{document}